\documentclass[12pt]{article}
\usepackage{amsfonts}
\usepackage{upgreek}

\newtheorem{proposition}{Proposition}

\begin{document}

\title{A discrete analogue of the modified Novikov-Veselov hierarchy}

\author{D.V.Zakharov \thanks{Columbia University, New York, USA; e-mail:
zakharov@math.columbia.edu}}

\date{}

\maketitle
\begin{abstract}
We construct a discrete analogue of the integrable two-dimensional Dirac operator and describe
the spectral properties of its eigenfunctions. We construct an integrable discrete analogue
of the modified Novikov-Veselov hierarchy. We derive the first two equations of the hierarchy and give explicit formulas for
the eigenfunctions in terms of the theta-functions of the associated spectral curve.

\end{abstract}

\section{Introduction}

\renewcommand{\theequation}{\thesection.\arabic{equation}}

\setcounter{equation}{0}

The purpose of this paper is to construct a discrete analogue of the modified
Novikov-Veselov hierarchy and its algebro-geometric solutions,
and to describe the spectral theory of the corresponding
discrete Dirac operator.

The modified Novikov--Veselov (mNV) hierarchy is an integrable hierarchy of equations introduced
by Bogdanov in \cite{bogdanov1}, \cite{bogdanov2} as a special reduction of the Davey--Stewardson equation.
The equations of the hierarchy have the form of Manakov $L,A,B$-triples
\begin{equation}
\frac{\partial L}{\partial t_n}=[L,A_n]-B_nL,
\end{equation}
where $L=D$ is the two-dimensional Dirac operator
\begin{equation}
\label{dir}
D\psi=\left(\begin{array}{cc}
u & \partial \\
-\bar\partial & u
\end{array}\right)
\left(\begin{array}{c}
\psi_1\\
\psi_2\end{array}\right),
\end{equation}
and $A_n$ and $B_n$ are $(2\times 2)$-matrix differential operators. The mNV hierarchy describes deformations
of the Dirac operator that preserve the zero energy level, i.e. isospectral deformations of the equation
\begin{equation}
D\psi=0.
\end{equation}
The first equation of the hierarchy has the form
\begin{equation}
\label{mNV}
u_t=\left(u_{zzz}+3u_zv+\frac{3}{2}uv_z\right)+
\left(u_{\bar z\bar z\bar z}+3u_{\bar z}v+\frac{3}{2}uv_{\bar z}\right),\hspace{4mm} v_{\bar z}=(u^2)_z.
\end{equation}
In \cite{taimanov1}, \cite{taimanov2} Taimanov constructed algebro-geometric solutions of the mNV hierarchy
and described the spectral theory of the Dirac operator (\ref{dir}). 
In recent times, the mNV hierarchy and its algebro-geometric solutions have
attracted significant attention due to their applications
to the classical theory of two-dimensional surfaces in three-dimensional Euclidean space, and in particular
to the Willmore conjecture (see the survey \cite{taimanov3} for an extensive bibliography). 

It is possible to consider a more general two-dimensional Dirac operator of the form
\begin{equation}
\label{dir1}
D=\left(\begin{array}{cc}
u & \partial \\
-\bar{\partial}& v
\end{array}\right).
\end{equation}
The spectral theory of the two-dimensional Dirac operator (\ref{dir1}) is equivalent to that of
the two-dimensional scalar Schr\"odinger operator in a magnetic field 
\begin{equation}
\label{schrodinger-magn}
H=\partial\bar{\partial} +V\bar{\partial}+U.
\end{equation}
The reduction of the Dirac operator (\ref{dir1}) to the form (\ref{dir}) corresponds to a reduction
on the Schr\"odinger operator in which the functions $U$ and $V$ satisty the relation
\begin{equation}
\label{red}
V=-\partial\ln U.
\end{equation}
The analytic properties of Baker--Akhiezer functions which describe general Schr\"odinger operators of 
the form (\ref{schrodinger-magn}) that are integrable on the zero energy level were formulated in \cite{dubrovin-krichever-novikov}. The reductions on the algebro-geometric data that describe the potential Schr\"odinger operator ($V=0$),
which is the auxiliary operator for the Novikov--Veselov hierarchy,
were found in \cite{novikov-veselov1}, \cite{novikov-veselov2}.

The problem of constructing an integrable discretization of an integrable
differential equation is not well-posed and does not have a universal solution. However,
there are several methods in soliton theory that allow us to construct integrable
discretizations. Most of them are based on constucting a discrete analogue of 
the auxiliary linear problems, which involves an appropriate deformation of the analytic
properties of the solutions of these linear problems. 

In the finite-gap case, the eigenfunction of the auxiliary linear differential operator, known as the
Baker--Akhiezer function, is defined
on an algebraic Riemann surface and is required to have exponential singularities controlled
by the continuous variables at one or
more marked points of the surface. To construct a discrete analogue of the operator, we replace
each exponential singularity with a pair of meromorphic singularities consising of a pole and a zero
of the same order, which we consider as the discrete variable. This deformed eigenfunction then
satisfies a infinite system of linear difference and differential equations, whose compatibility conditions
are the discretization of the original integrable hierarchy.
This method was used for constructing algebro-geometric solutions of the Ablowitz--Ladik equation \cite{ablowitz-ladik}, \cite{miller-ercolani-krichever-levermore},
which is a discretization of the nonlinear Schr\"odinger
equation, and for constructing Darboux--Egoroff lattices, which are the discrete analogue
of Darboux--Egoroff metrics \cite{discrete-egoroff}.

Using this approach, Grushevsky and Krichever have given an algebro-geometric
construction of an integrable discretization of the two-dimensional Schr\"odinger
operator (\ref{schrodinger-magn}). In the second paragraph, we describe a matrix variant
of this construction, which leads to a two-dimensional matrix difference operator of the form
\begin{equation}
\label{ddir}
D{\bf \psi}=\left[\left(\begin{array}{cc} T_2 & 0 \\
0 & T_1\end{array}\right)-
\left(\begin{array}{cc}\alpha & \beta \\ \gamma & \delta \end{array}\right)
\right]\left(\begin{array}{c}\psi_1 \\ \psi_2\end{array}\right),
\end{equation}
where the functions $\psi_i$ and the coefficients of the operator are functions of two discrete
variables $n,m\in\mathbb{Z}$, and $T_1, T_2$ denote the translation operators in the
discrete variables. The operator $D$, which we call the {\it discrete Dirac operator}, can be
considered as a discrete analogue of the general Dirac operator of the form $(\ref{dir1})$.

The coefficients of a discrete Dirac operator (\ref{ddir}) depend, up to gauge transformation,
on two arbitrary functions of the discrete variables. In the second paragraph, we show that
a discretization of the algebro-geometric data corresponding to operators of the form (\ref{dir})
leads to operators whose coefficients depend on only one arbitrary function, namely operators of
the form
\begin{equation}
\label{Ddir}
D\psi=
\left[\left(\begin{array}{cc} T_2 & 0 \\
0 & T_1\end{array}\right)-
\left(\begin{array}{cc}\alpha & \beta \\ \beta & \alpha \end{array}\right)
\right]\left(\begin{array}{c}\psi_1 \\ \psi_2\end{array}\right),
\end{equation}
where the coefficients satisfy the relation
\begin{equation}
\label{const}
\alpha^2-\beta^2=1
\end{equation}
In the third paragraph we introduce time dependence into the eigenfunctions and construct an integrable hierarchy of isospectral deformations of the zero energy level of the operator (\ref{Ddir}). We call this hierarchy,
which has the form of Manakov $L,A,B$-triples, the {\it discrete modified Novikov-Veselov hierarchy}.

In the fourth paragraph we derive the explicit form of the first two equations of
the hierarchy (equations (\ref{t11final}), (\ref{t12final}), (\ref{t21final})). The first equation has the following form:
\begin{equation}
\frac{\partial \psi(n,m)}{\partial \tau_1^1}=\sqrt{\left(e^{2 \varphi(n-1,m+1)} -e^{2\varphi(n-1,m)}\right)
\left(e^{-2\varphi(n,m)}-e^{-2\varphi(n,m+1)}\right)},
\end{equation}
where the two functions satisfy the non-local relation
\begin{equation}
\varphi(n,m+1)-\varphi(n,m)=\psi(n+1,m)-\psi(n,m).
\end{equation}
In the final paragraph we give explicit formulas for the Baker--Akhiezer functions in terms of theta-functions associated to the spectral curve.

\section{Reduction of general discrete Dirac operators}

\setcounter{equation}{0}

Consider the following discrete linear equation
\begin{equation}
D{\bf \psi}=\left[\left(\begin{array}{cc} T_2 & 0 \\
0 & T_1\end{array}\right)-
\left(\begin{array}{cc}\alpha & \beta \\ \gamma & \delta \end{array}\right)
\right]\left(\begin{array}{c}\psi_1 \\ \psi_2\end{array}\right)=0,
\label{Dirac}
\end{equation}
where $\psi=(\psi_1(n,m), \psi_2(n,m))^T$ is a vector function of two discrete variables $n,m\in \mathbb{Z}$,
and
\begin{equation}
\left(\begin{array}{cc}\alpha & \beta \\ \gamma & \delta \end{array}\right)
=\left(\begin{array}{cc}\alpha(n,m) & \beta(n,m)\\ \gamma(n,m) & \delta(n,m)\end{array} \right)
\end{equation}
is a $(2\times2)$-matrix function of the discrete variables. We call $D$ the {\it discrete Dirac operator}.
 We use $T_1$ and $T_2$ to denote the 
translation operators in the discrete variables
\begin{equation}
T_1f(n,m)=f(n+1,m),\hspace{4mm} T_2f(n,m)=f(n,m+1),
\end{equation}
while $t_1$ and $t_2$ will be used to denote the translated functions, so that for example $T_1(fg)=(t_1f)(t_1g)$. In this
chapter, we construct algebro-geometric solutions of equation $(\ref{Dirac})$ and some of its reductions.

The main method of constructing algebro-geometric solutions of linear differential or difference equations
such as (\ref{Dirac}) is to consider functions $\psi_i$ defined on an auxiliary Riemann surface, called the {\it spectral
curve}, and having certain prescribed singularities on that curve. Generally, to construct solutions of difference equations, we consider functions that are meromorphic on the spectral curve with prescribed pole singularities, while constructing
solutions of differential equations requires us to consider functions with prescribed essential singularities, called {\it
Baker-Akhiezer functions}. 
 
Let $X$ be a smooth Riemann surface of genus $g$. We consider the following data on X:

\noindent {\bf Data A.}
\begin{itemize} \item Four distinct marked points $P_1^{\pm}, P_2^{\pm}$ on $X$.

\item Local parameters $z_i^{\pm}=(k_i^{\pm})^{-1}$ defined in some neighborhoods of these points.

\item An effective divisor $D=\gamma_1+\cdots+\gamma_{g+1}$ of degree $g+1$ on $X$, supported away
from the marked points, which satisfies the following condition of general position:
\begin{equation}
h^1(D+(n-1)P_1^+-nP_1^-+(m-1)P_2^+-mP_2^-)=0\mbox{ for all }n,m\in\mathbb{Z}.
\label{Disgeneral}
\end{equation}

\end{itemize}

To construct solutions of equation (\ref{Dirac}), we consider spaces of meromorphic functions on $X$ with singularities controlled by the discrete variables:
$$
\Psi_{n,m}=H^0(D+nP_1^+-nP_1^-+mP_2^+-mP_2^-)\subset \mbox{Mer}(X),\hspace{4mm}n,m\in\mathbb{Z}.
$$
The Riemann-Roch theorem implies the following
\begin{proposition}
Suppose that $X$ is an algebraic curve with data A defined above. Then each of the spaces $\Psi_{n,m}$ is two-dimensional:
$$
\dim \Psi_{n,m}=h^0(D+nP_1^+-nP_1^-+mP_2^+-mP_2^-)=2\mbox{ for all }n,m\in\mathbb{Z},
$$
the intersection of two of these spaces at adjacent lattice points is one-dimensional:
$$
\dim \Psi_{n,m}\cap \Psi_{n,m-1}=h^0(D+nP_1^+-nP_1^-+(m-1)P_2^+-mP_2^-)=1\mbox{ for all }n,m\in\mathbb{Z},
$$
$$
\dim \Psi_{n,m}\cap \Psi_{n-1,m}=h^0(D+(n-1)P_1^+-nP_1^-+mP_2^+-mP_2^-)=1\mbox{ for all }n,m\in\mathbb{Z},
$$
and these two one-dimensional subspaces of $\Psi_{n,m}$ span the entire space, i.e. their intersection is trivial:
$$
\dim \Psi_{n,m}\cap\Psi_{n,m-1}\cap \Psi_{n-1,m}=h^0(D+(n-1)P_1^+-nP_1^-+(m-1)P_2^+-mP_2^-)=0\mbox{ for all }n,m\in\mathbb{Z}.
$$
\end{proposition}

Therefore, we can fix a basis $\psi_1(n,m,P), \psi_2(n,m,P)$ in each of the spaces
$\Psi_{n,m}$ by letting $\psi_1(n,m,P)$ be any non-zero element of $\Psi_{n,m}\cap \Psi_{n,m-1}$, 
and letting $\psi_2(n,m,P)$ to be any non-zero element of $\Psi_{n,m}\cap \Psi_{n-1,m}$:
\begin{equation}
\psi_1(n,m,P)\in H^0(D+nP_1^+-nP_1^-+(m-1)P_2^+-mP_2^-)-\{0\},
\label{psi1}
\end{equation}
\begin{equation}
\psi_2(n,m,P)\in H^0(D+(n-1)P_1^+-nP_1^-+mP_2^+-mP_2^-)-\{0\}.
\label{psi2}
\end{equation}

The principal observation concerning these functions can be summarized in the following statement:
\begin{proposition}
Suppose that $X$ is a Riemann surface with data A as defined above. Then there exist functions $\alpha(n,m)$, $\beta(n,m)$, $\gamma(n,m)$, $\delta(n,m)$ such that the functions $\psi_1(n,m,P)$
and $\psi_2(n,m,P)$ defined by (\ref{psi1})-(\ref{psi2}) satisfy the Dirac equation:
\begin{equation}
D\psi=\left[\left(\begin{array}{cc} T_2 & 0 \\
0 & T_1\end{array}\right)-
\left(\begin{array}{cc}\alpha(n,m) & \beta(n,m) \\ \gamma(n,m) & \delta(n,m) \end{array}\right)
\right]\left(\begin{array}{c}\psi_1(n,m,P) \\ \psi_2(n,m,P)\end{array}\right)=0.
\label{Dirac2}
\end{equation}\end{proposition}

\noindent {\bf Proof. }Indeed, by construction, both $\psi_1(n,m+1,P)$ and $\psi_2(n+1,m,P)$ 
actually lie in the space $\Psi_{n,m}$, hence they can be expressed as linear combinations of the basis functions $\psi_1(n,m,P)$ and $\psi_2(n,m,P)$, which is equivalent to saying that the satisfy the Dirac equation (\ref{Dirac2}).

Therefore, a Riemann surface $X$ together with the additional data given above allows us to construct a family
of solutions $(\psi_1(n,m,P),\psi_2(n,m,P))^T$ of the Dirac equation (\ref{Dirac}), parametrized by the points $P$
of $X$.

In order to construct reductions on the Dirac equation (\ref{Dirac2}), we first express the coefficients $\alpha(n,m)$,
$\beta(n,m)$, $\gamma(n,m)$ and $\delta(n,m)$ in terms of the principal parts of the basis functions at the marked
points. In terms of the chosen local coordinates, the basis functions $\psi_1(n,m,P)$ and $\psi_2(n,m,P)$ have the following expansions at the marked points, where $k$ denotes the appropriate local parameter $k_{\pm}^i$:
\begin{equation}
\psi_1(n,m,P)=\left\{\begin{array}{cc}a_1^+(n,m)k^n+O(k^{n-1}), &\mbox{as }P\rightarrow P_1^+\\
a_1^-(n,m)k^{-n}+O(k^{-n-1}), &\mbox{as }P\rightarrow  P_1^-\\
O(k^{m-1}), &\mbox{as }P\rightarrow  P_2^+\\
a_2^-(n,m)k^{-m}+O(k^{-m-1}), &\mbox{as }P\rightarrow  P_2^-\end{array}\right.
\label{psi1expansionsnotime}
\end{equation}
\begin{equation}
\psi_2(n,m,P)=\left\{\begin{array}{cc}O(k^{n-1}), &\mbox{as }P\rightarrow  P_+^1\\
b_1^-(n,m)k^{-n}+O(k^{-n-1}), &\mbox{as }P\rightarrow  P_-^1\\
b_2^+(n,m)k^{m}+O(k^{m-1}), &\mbox{as }P\rightarrow  P_+^2\\
b_2^-(n,m)k^{-m}+O(k^{-m-1}), &\mbox{as }P\rightarrow  P_-^2\end{array}\right.
\label{psi2expansionsnotime}
\end{equation}
where the $a_i^{\pm}(n,m)$ and $b_i^{\pm}(n,m)$ are functions of the discrete variables $n$ and $m$. Considering
the Dirac equation (\ref{Dirac2}) near the marked points $P_1^{\pm}, P_2^{\pm}$ gives us the following system
of equations (in what follows, we usually suppress the indices $n$ and $m$ and replace them with the shift operators $t_1$ and $t_2$):
\begin{equation}
\begin{array}{ccc}
t_2a_1^+ & = & \alpha a_1^+,\\
t_2a_1^- & = & \alpha a_1^-+\beta b_1^-,\\
0 & = & \alpha a_2^-+\beta a_2^+,\end{array}\hspace{4mm}
\begin{array}{ccc}
0 & = & \gamma a_1^-+\delta b_1^-,\\
t_1b_2^+ & = & \delta b_2^+,\\
t_1b_2^- & = & \gamma a_2^-+\delta b_2^-.\end{array}
\label{Eight}
\end{equation}

The functions $\psi_1$ and $\psi_2$ have so far been defined up to multiplication by a constant factor dependent
on $n$ and $m$. We impose the
following additional condition on the functions $\psi_1$ and $\psi_2$:
\begin{equation}
a_1^+a_1^-=1,\hspace{10mm}b_2^+b_2^-=1.
\label{Gauge1}
\end{equation}
It is easy to show using (\ref{Eight}) that these conditions imply the following relations on the
coefficients $\alpha$, $\beta$, $\gamma$, $\delta$:
\begin{equation}
\alpha\delta-\beta\gamma=\frac{\alpha}{\delta}=\frac{\delta}{\alpha}=\frac{(t_2a_1^+)(t_1b_2^-)}{a_1^+b_2^-}=\pm 1.
\label{Reduction1}
\end{equation}
Condition (\ref{Gauge1}) defines the constants $a_1^+$ and $b_2^-$, and hence the functions $\psi_1$ and $\psi_2$, only up to a factor of $\pm 1$ that depends on $n$ and $m$. This allows us to impose the following additional condition on the functions $\psi_1$ and
$\psi_2$:
\begin{equation}
(t_2a_1^+)(t_1b_2^-)=a_1^+b_2^-.
\label{Gauge2}
\end{equation}
In other words, we can choose the sign for the function $\psi_2$ arbitrarily, and then choose the sign for the
function $\psi_1$ using the above relation. With this condition, the sign in equation (\ref{Reduction1}) is positive. Therefore, reductions (\ref{Gauge1}) and
(\ref{Gauge2}) impose the following relations on the coefficients of the Dirac operator (\ref{Dirac2}):
\begin{equation}
\alpha\delta-\beta\gamma=1,\hspace{10mm}\alpha=\delta
\end{equation}
In other words, the coefficients of a general Dirac operator of the form (\ref{Dirac2}) depend, up to gauge
equivalence, on two arbitrary functions of the discrete variables.

We now introduce a reduction under which the coefficients of the Dirac operator (\ref{Dirac2}) depend on only
one function of the variables $n$, $m$. Suppose that, in addition to data A described
above, the spectral curve $X$ has the following

\noindent {\bf Data B.}
\begin{itemize}

\item A holomorphic involution $\sigma\!:\! X\rightarrow X$ that interchanges the marked points and the local
parameters at the marked points as follows:
\begin{equation}
\sigma(P_i^{\pm})=P_i^{\mp},\hspace{4mm}\sigma(k_i^{\pm})=k_i^{\mp}.
\end{equation}

\item A meromorphic $1$-form $\omega$ on $X$ which has simple poles at the marked points $P_i^{\pm}$ with residues $\pm 1$ and no other singularities, whose zero divisor is $D+\sigma(D)$, and which is odd with respect to the involution.
\end{itemize}
Consider the meromorphic $1$-form $\psi_1(n,m,P)\psi_2(n,m,\sigma(P))\omega(P)$. Comparing the singularities of the three
terms, we see that this $1$-form has simple poles at $P_1^+$ and $P_2^-$ with residues $a_1^+b_1^-$ and $-a_2^-b_2^+$,
respectively, and no other singularities. Hence, the existence of the additional data above implies that
the coefficients of the functions $\psi_1$ and $\psi_2$ satisfy the following additional condition:
\begin{equation}
a_1^+b_1^-=a_2^-b_2^+.
\end{equation}
Using (\ref{Eight}) and (\ref{Gauge1}), it is easy to show that this condition implies the following 
additional relation on the coefficients of the Dirac operator:
\begin{equation}
\beta=\gamma.
\end{equation}
Using the involution $\sigma$ we can rewrite the normalization conditions (\ref{Gauge1}) and (\ref{Gauge2}) 
in the following equivalent form:
\begin{equation}
\left.\psi_1(P)\psi_1(\sigma(P))\right|_{P=P_1^+}=1,
\label{norm1}
\end{equation}
\begin{equation}
\left.\psi_2(P)\psi_2(\sigma(P))\right|_{P=P_2^+}=1,
\label{norm2}
\end{equation}
\begin{equation}
\left.\frac{t_2\psi_1(P)}{\psi_1(P)}\right|_{P=P_1^+}=\left.\frac{t_1\psi_2(P)}{\psi_2(P)}\right|_{P=P_2^-}.
\label{norm3}
\end{equation}
Therefore, we can summarize the result of this reduction as follows.
\begin{proposition}
Suppose that $X$ is a Riemann surface with data A and data B as defined above, and suppose the functions
$\psi_1(P)$ and $\psi_2(P)$ defined by (\ref{psi1}) and (\ref{psi2}) satisfy the normalization conditions
(\ref{norm1})-(\ref{norm3}). Then there exist functions of the discrete variables $\alpha$ and $\beta$
that satisfy the relation
\begin{equation}
\alpha^2-\beta^2=1
\end{equation}
and such that the functions $\psi_1(P)$
and $\psi_2(P)$ satisfy the discrete Dirac equation:
\begin{equation}
D\psi=
\left[\left(\begin{array}{cc} T_2 & 0 \\
0 & T_1\end{array}\right)-
\left(\begin{array}{cc}\alpha & \beta \\ \beta & \alpha \end{array}\right)
\right]\left(\begin{array}{c}\psi_1(P) \\ \psi_2(P)\end{array}\right)=0.
\label{Dirac3}
\end{equation}\end{proposition}

We now construct a further reduction of the discrete Dirac equation (\ref{Dirac3}) which is the discrete analogue
of the real-valued reduction in the differential case. Suppose that, in addition to data A and data B above, the 
spectral curve $X$ has the following 

\noindent {\bf Data C}.
\begin{itemize}
\item An anti-holomorphic involution $\tau\!:\! X\rightarrow X$ that interchanges the marked points and
acts on the local parameters at the marked points as follows:
\begin{equation}
\tau(P_1^{\pm})=P_2^{\pm},\hspace{4mm}\tau(P_2^{\pm})=P_1^{\pm},\hspace{4mm}
\tau(k_1^{\pm})=\bar{k}_2^{\pm},\hspace{4mm}\tau(k_2^{\pm})=\bar{k}_1^{\pm}.
\end{equation}
\item A meromorphic function $f(P)$ on $X$ with divisor $(f)=D-\tau(D)$ satisfying the conditions
\begin{equation}
f(P)\bar{f}(\tau(P))=-1\mbox{ for all }P\in X,\hspace{4mm}f(P_1^+)f(P_1^-)=1.
\label{realreduction}
\end{equation}
\end{itemize}

For a function $f(n,m)$ of the discrete variables, we introduce the notation $f^*(n,m)=\bar{f}(m,n)$.
Consider the two functions $\psi_2^*(n,m,\tau(P))$ and $\psi_1(n,m,P)f(P)$. Both these
functions are meromorphic and lie in the one-dimensional space $H^0(\tau(D)+(n-1)P_2^+-nP_2^-+mP_1^+-mP_1^-)$, hence
there exists a function $C(n,m)$ of $n$ and $m$ such that
\begin{equation}
\bar{\psi}_2(m,n,\tau(P))=\psi_1(n,m,P)f(P)C(n,m).
\end{equation}
Considering this equation at $P=P_1^+$ and $P=P_1^-$ and using conditions (\ref{Gauge1}) and (\ref{realreduction}), we
see that
\begin{equation}
C(n,m)^2=1\mbox{ for all }n,m\in\mathbb{Z}.
\end{equation}
We recall that the function $\psi_2$ was normalized by condition (\ref{Gauge1}), which specifies it up to multiplication
by a factor $\pm 1$ dependent on $n$ and $m$. Therefore, we can choose this factor in such a way that $C(n,m)=1$ for all
$n$ and $m$, in other words we may impose the additional following condition:
\begin{equation}
\bar{\psi}_2(m,n,\tau(P))=\psi_1(n,m,P)f(P).
\label{norm4}
\end{equation}
Equation (\ref{realreduction}) then implies that the functions $\psi_1$ and $\psi_2$ chosen in this way satisfy the following relations:
\begin{equation}
\bar{\psi}_2(m,n,\tau(P))=\psi_1(n,m,P)f(P),\hspace{4mm}\bar{\psi}_1(m,n,\tau(P))=-\psi_2(n,m,P)f(P).
\end{equation}
Plugging these relations into the reduced Dirac equation (\ref{Dirac3}) gives us the following relations on 
the coefficients of the operator:
\begin{equation}
\alpha^*=\alpha,\hspace{4mm}\beta^*=-\beta.
\end{equation}
We summarize the results of this reduction in the following proposition:
\begin{proposition}
Suppose that $X$ is an algebraic curve with data A, B and C as defined above, and suppose the functions
$\psi_1(P)$ and $\psi_2(P)$ defined by (\ref{psi1}) and (\ref{psi2}) satisfy the normalization conditions
(\ref{norm1})-(\ref{norm3}) and (\ref{norm4}). Then there exist functions of the discrete
variables $\alpha$ and $\beta$ that satisfy the relations
\begin{equation}
\alpha^2-\beta^2=1,\hspace{4mm}\alpha^*=\alpha,\hspace{4mm}\beta^*=-\beta
\end{equation}
that the functions $\psi_1(P)$
and $\psi_2(P)$ satisfy the discrete Dirac equation:
\begin{equation}
D\psi=
\left[\left(\begin{array}{cc} T_2 & 0 \\
0 & T_1\end{array}\right)-
\left(\begin{array}{cc}\alpha & \beta \\ \beta & \alpha \end{array}\right)
\right]\left(\begin{array}{c}\psi_1(P) \\ \psi_2(P)\end{array}\right)=0.
\label{Dirac4}
\end{equation}\end{proposition}

\section{The discrete modified Novikov-Veselov hierarchy}

\setcounter{equation}{0}

In the previous section, we constructed algebro-geometric solutions of the discrete Dirac operator (\ref{Dirac2}) and
its reductions (\ref{Dirac3}) and (\ref{Dirac4}) by considering spaces of meromorphic functions $\Psi_{n,m}$ on
an algebraic curve $X$ with poles and zeroes determined by the numbers $n$ and $m$. In this section, we embed these meromorphic solutions into a family of transcendental functions, called {\it Baker-Akhiezer functions}, and construct a hierarchy of commuting flows on the space of these functions. The set of compatibility conditions of these flows is the discrete
analogue of the modified Novikov-Veselov hierarchy.

Let $\tau=\left\{\tau_s^1,\tau_s^2, s=1,2,\ldots\right\}\in\mathbb{C}^{\infty}\oplus\mathbb{C}^{\infty}$ denote two sequences of complex numbers, only finitely many of
which are non-zero, which we think of as continuous time variables. 
We construct deformations $\Psi_{n,m,\tau}$ of the function spaces $\Psi_{n,m}$ constructed in Section 2 by
considering functions which in addition have essential singularities at the marked points controlled by the times $\tau$.

\begin{proposition}
Suppose that $X$ is an algebraic curve with data A and data B given as in the previous section. Denote
by $\tilde{X}=X-P_1^+-P_1^--P_2^+-P_2^-$ the curve $X$ with the marked points removed. Consider
the space $\Psi_{n,m,\tau}\in\mbox{Mer}(\tilde{X})$ of functions on $\tilde{X}$ defined by the following conditions
\begin{enumerate}
\item For all $\psi(n,m,\tau;P)\in\Psi_{n,m,\tau}$ we have $(\psi)+D\geq 0$, where $(f)$ denotes the divisor of $f$.

\item At the marked points $P_i^{\pm}$ the elements $\psi(n,m,\tau;P)$ of $\Psi_{n,m,\tau}$ have essential
singularities of the following form, where by $k$ we denote the appropriate local coordinate $k_i^{\pm}$:
\begin{equation}
\begin{array}{c}
\psi(n,m,\tau;P)=\exp\left(\pm\displaystyle\sum_{s=1}^{\infty}\tau_s^1k^s\right)
O(k^{\pm n})\mbox{ as }P\rightarrow P_1^{\pm},\\
\psi(n,m,\tau;P)=\exp\left(\pm\displaystyle\sum_{s=1}^{\infty}\tau_s^2k^s\right)
O(k^{\pm m})\mbox{ as }P\rightarrow P_2^{\pm}.\end{array}
\end{equation}
\end{enumerate}
Then each of the spaces $\Psi_{n,m,\tau}$ is two-dimensional:
\begin{equation}
\dim \Psi_{n,m,\tau}=2\mbox{ for all }n,m\in\mathbb{Z},
\end{equation}
the intersection of two of these spaces at adjacent lattice points is one-dimensional:
\begin{equation}
\dim \Psi_{n,m,\tau}\cap \Psi_{n,m-1,\tau}=1\mbox{ for all }n,m\in\mathbb{Z},
\end{equation}
\begin{equation}
\dim \Psi_{n,m,\tau}\cap \Psi_{n-1,m,\tau}=1\mbox{ for all }n,m\in\mathbb{Z},
\end{equation}
and these two one-dimensional subspaces of $\Psi_{n,m,\tau}$ span the entire space, i.e. intersection is trivial:
\begin{equation}
\dim \Psi_{n,m,\tau}\cap\Psi_{n,m-1,\tau}\cap \Psi_{n-1,m,\tau}=0\mbox{ for all }n,m\in\mathbb{Z}.
\label{zero}
\end{equation}
\end{proposition}

\noindent {\bf Proof.} The proof of this proposition is a standard application of the Riemann--Roch theorem.

This proposition allows us to define functions $\psi_1(n,m,\tau;P)$ and $\psi_2(n,m,\tau;P)$ using the same
relations as in Section 2. We observe the normalization conditions (\ref{norm1})-(\ref{norm3}) can be applied
to elements of $\Psi_{n,m,\tau}$, since the exponential singularities cancel out.

\begin{proposition}  There exist unique functions $\psi_1(n,m,\tau;P)$ and $\psi_2(n,m,\tau;P)$ that
form a basis for the vector space $\Psi_{n,m,\tau}$ such that
\begin{equation}
\psi_1(n,m,\tau;P)\in \Psi_{n,m,\tau}\cap \Psi_{n,m-1,\tau}-\{0\},
\label{psi1time}
\end{equation}
\begin{equation}
\psi_2(n,m,\tau;P)\in \Psi_{n,m,\tau}\cap \Psi_{n-1,m,\tau}-\{0\}.
\label{psi2time}
\end{equation}
and which satisfy the normalization conditions (\ref{norm1})-(\ref{norm3}). These functions satisfy the discrete Dirac equation
\begin{equation}
D\psi=
\left[\left(\begin{array}{cc} T_2 & 0 \\
0 & T_1\end{array}\right)-
\left(\begin{array}{cc}\alpha & \beta \\ \beta & \alpha \end{array}\right)
\right]\left(\begin{array}{c}\psi_1 \\ \psi_2\end{array}\right)=0,
\label{Diractime}
\end{equation}
where $\alpha$ and $\beta$ are functions of the variables $n$, $m$, and ${\bf \tau}$ that satisfy the condition
\begin{equation}
\alpha^2-\beta^2=1.
\end{equation}
\end{proposition}
In Section 6, we give explicit formulas for the functions $\psi_i$ in terms of theta-functions.

We now show that these functions satisfy a system of commuting linear equations. Let $\mathfrak{R}$ denote the ring of functions in the variables $n$, $m$ and $\tau$. We consider the ring
$\mathfrak{O}=\mathfrak{R}[T_1,T_1^{-1},T_2,T_2^{-1}]$ of finite difference operators with coefficients in $\mathfrak{R}$,
and the ring $\mathfrak{M}$ of $(2\times2)$ matrix operators with coefficients in $\mathfrak{O}$. By $\psi$ we denote the
column vector $(\psi_1(n,m,\tau;P,\psi_2(n,m,\tau;P))^T$.

\begin{proposition}
There exist unique matrix difference operators $A_{s}^{i}$ in $\mathfrak{M}$
\begin{equation}
A_s^i=\left(\begin{array}{cc}A_{s,1}^i & 0\\ 0 & A_{s,2}^i\end{array}\right),\hspace{4mm}i=1,2,
\end{equation}
\begin{equation}
A_{s,j}^{i}=\displaystyle\sum_{\mu=-s}^{s}f_{s,j,\mu}^{i}(n,m,\tau)T_i^{\mu},
\end{equation}
such that the functions $\psi_1(n,m,\tau;P)$
and $\psi_2(n,m,\tau;P)$ satisfy the following system of differential equations:
\begin{equation}
\frac{\partial }{\partial \tau_s^{i}}\psi=A^{i}_{s}\psi.
\label{psitime}
\end{equation}
\end{proposition}

\noindent {\bf Proof.} The proof is standard. For a given $s$ we show how to construct the operator $A_{s,1}^1$, the other cases being similar.

The derivative of the function $\psi_1(n,m,\tau;P)$ with respect to $\tau_s^1$ has the following expansions at the marked points $P_i^{\pm}$, where by $k$ we denote the appropriate local coordinate $k_i^{\pm}$:
\begin{equation}
\frac{\partial }{\partial \tau_s^1}\psi_1(n,m,\tau;P)=\exp\left(\pm\displaystyle\sum_{\sigma=1}^{\infty}\tau_{\sigma}^1k^{\sigma}\right)\cdot
O(k^{\pm n+s})\mbox{ as }P\rightarrow P_1^{\pm},
\end{equation}
\begin{equation}
\frac{\partial }{\partial \tau_s^1}\psi_1(n,m,\tau;P)=\exp\left(\displaystyle\sum_{\sigma=1}^{\infty}\tau_{\sigma}^2k^{\sigma}\right)O(k^{m-1})
\mbox{ as }P\rightarrow P_2^+,
\label{p2plus}
\end{equation}
\begin{equation}
\frac{\partial }{\partial \tau_s^1}\psi_1(n,m,\tau;P)
=\exp\left(-\displaystyle\sum_{\sigma=1}^{\infty}\tau_{\sigma}^2k^{\sigma}\right)\cdot
O(k^{-m})\mbox{ as }P\rightarrow P_2^-.
\label{p2minus}
\end{equation}
Therefore, for an appropriate choice of functions $f^1_{s,i,\mu}(n,m,\tau)$, the function
\begin{equation}
\tilde{\psi}(n,m,\tau;P)=\frac{\partial }{\partial \tau_s^1}\psi_1(n,m,\tau;P)-
\displaystyle\sum_{\mu=-s}^{s}f_{s,1,\mu}^{1}(n,m,\tau)\psi_1(n+\mu,m,\tau;P)
\end{equation}
has the following expansions at $P_1^{\pm}$:
\begin{equation}
\tilde{\psi}(n,m,\tau;P)=\exp\left(\pm\displaystyle\sum_{\sigma=1}^{\infty}\tau_{\sigma}^1k^{\sigma}\right)\cdot
O(k^{n-1})\mbox{ as }P\rightarrow P_1^{+},
\end{equation}
\begin{equation}
\tilde{\psi}(n,m,\tau;P)=\exp\left(\pm\displaystyle\sum_{\sigma=1}^{\infty}\tau_{\sigma}^1k^{\sigma}\right)\cdot
O(k^{-n})\mbox{ as }P\rightarrow P_1^{-},
\end{equation}
and the same expansions (\ref{p2plus})-(\ref{p2minus}) at $P_2^{\pm}$
 as $\frac{\partial }{\partial \tau_s^1}\psi_1(n,m,\tau;P)$. Therefore,
by (\ref{zero}) this function is identically zero on $X$. Hence, the function $\psi_1(n,m,\tau;P)$ satisfies
the system of equations (\ref{psitime}).

\begin{proposition}
The left ideal of matrix difference operators in $\mathfrak{M}$ that annihilate $\psi$ is the principal left ideal generated by the operator $D$.
\end{proposition}

\noindent {\bf Proof.} Suppose that $A$ and $B$ are two operators in $\mathfrak{O}$ that satisfy the following equation:
\begin{equation}
A\psi_1+B\psi_2=0.
\label{ideal}
\end{equation}
We need to show that there exist elements $C,D\in\mathfrak{O}$ such that $A=C(T_2-\alpha)-D\beta$ and
$B=-C\beta+D(T_1-\alpha)$.

First, we multiply equation (\ref{ideal}) on the left by sufficiently high powers of $T_1$ and $T_2$ so that the operators $A$ and $B$ become polynomial in $T_1$ and $T_2$. Next, we show that we can eliminate all terms containing mixed
powers of $T_1$ and $T_2$. Indeed, suppose
$$
A=\displaystyle\sum_{i=1}^{n-1}a_{i}T_1^iT_2^{n-i}+(\mbox{terms with no $T_1T_2$})+
(\mbox{terms of order $<n$}),
$$
$$
B=\displaystyle\sum_{i=1}^{n-1}b_{i}T_1^iT_2^{n-i}+(\mbox{terms with no $T_1T_2$})+
(\mbox{terms of order $<n$}),
$$
then we can write
$$
A=\displaystyle\sum_{i=1}^{n-1}\left[a_iT_1^iT_2^{n-i-1}(T_2-\alpha)-b_iT_1^iT_2^{n-i-1}\beta\right]+(\mbox{terms with no
$T_1T_2$})+(\mbox{terms of order $<n$}),
$$
$$
B=\displaystyle\sum_{i=1}^{n-1}\left[b_iT_1^iT_2^{n-i-1}(T_1-\alpha)-a_iT_1^iT_2^{n-i-1}\alpha\right]+(\mbox{terms with no
$T_1T_2$})+(\mbox{terms of order $<n$}),
$$
and proceeding in this way, we can eliminate all terms which are not powers of only $T_1$ or $T_2$. Therefore, we can assume
that $A=A_1(T_1)+A_2(T_2)$, $B=B_1(T_1)+B_2(T_2)$, where the $A_i, B_i$ are polynomials in only $T_i$.

Suppose that $A_1=\sum_{i=0}^n a_i T_1^i$ and $B_1=\sum_{j=0}^m b_jT_1^j$. Comparing the singularities in (\ref{ideal}) at the
point $P_1^+$, we see that $m=n+1$. Subtracting $b_{n+1}T_1^n\left[(T_1-\alpha)\psi_2-\beta\psi_1\right]$ from (\ref{ideal}),
we reduce the degree of $B_1$, and hence of $A_1$. In this way we can eliminate $A_1$, and similarly $B_2$. Therefore, we
are left with showing that if $A=A_2(T_2)$ and $B=B_1(T_1)$ are linear polynomials satisfying (\ref{ideal}), then they can be expressed as $A=f(T_2-\alpha)-g\beta$ and $B=-f\beta+g(T_1-\alpha)$ for some functions $f$ and $g$, which can be easily shown.

\begin{proposition}
There exist matrix difference operators $B_s^i$ in $\mathfrak{M}$ such that the
following equations are satisfied:
\begin{equation}
-\frac{\partial}{\partial t_s^i}D=DA_s^i+B_s^iD
\end{equation}
\end{proposition}

\noindent {\bf Proof.} Equations (\ref{Diractime}) and (\ref{psitime}) imply that
\begin{equation}
\left[\frac{\partial}{\partial t_s^i}-A_s^i,D\right]\psi=0.
\end{equation}
Since the operator in the left hand side does not contain derivation in time, it is inside $\mathfrak{M}$, hence by the above
  proposition it is a left multiple of $D$, which proves the statement.

\begin{proposition}
The equations
\begin{equation}
\frac{\partial}{\partial t_s^i}D+DA_s^i\equiv 0 \mbox{ mod } D
\label{dmNV}
\end{equation}
define a commuting hierarchy of differential-difference equations.
\end{proposition}

We call this system the {\it discrete modified Novikov-Veselov (dmNV) hierarchy}. In the next section,
we give the explicit form of the first two pairs of equations of the dmNV hierarchy.

\section{First and second equations: explicit forms}

\setcounter{equation}{0}

In this section, we write down the explicit form of the dmNV hierarchy corresponding
to times $\tau_1^1$, $\tau_1^2$, $\tau_2^1$ and $\tau_2^2$. We give the explicit calculations for $\tau_1^1$, the derivations
for the other times being similar.

It is difficult to write down the dmNV as they are defined in (\ref{dmNV}), since 
this involves performing division with remainder in a matrix algebra over a non-commutative operator ring. To circumvent
this difficulty, we notice that the discrete Dirac equation (\ref{Diractime}), which is a difference equation of degree one
on the two functions $\psi_1$ and $\psi_2$, is equivalent to a degree two difference equation on one of the $\psi_1$ or $\psi_2$.

\begin{proposition}
Suppose the functions $\psi_1$ and $\psi_2$ satisfy the discrete Dirac equation (\ref{Diractime}). Then the functions $\psi_1$
and $\psi_2$ satisfy the following discrete Schr\"odinger equations
\begin{equation}
H_1\psi_1=\left[T_1T_2-(t_1\alpha)T_1-\frac{\alpha(t_1\beta)}{\beta}T_2+\frac{t_1\beta}{\beta}\right]\psi_1=0
\label{Sch}
\end{equation}
\begin{equation}
H_2\psi_2=\left[T_1T_2-(t_2\alpha)T_2-\frac{\alpha(t_2\beta)}{\beta}T_1+\frac{t_2\beta}{\beta}\right]\psi_2=0.
\end{equation}
\end{proposition}

\noindent {\bf Proof.} This follows from excluding $\psi_1$ or $\psi_2$ from the system (\ref{Diractime}).

Conversely, we have an analogue of Proposition 3.4 for the operators $H_i$:

\begin{proposition}
The left ideal of difference operators in $\mathfrak{O}$ that annihilate $\psi_i$ is the principal left ideal generated by the operator $H_i$.
\end{proposition}

\noindent {\bf Proof.} Suppose that $A\in\mathfrak{O}$ is an operator such that $A\psi_1=0$. Then Proposition 3.4 implies that
there exist operators $C,D\in\mathfrak{O}$ such that
$$
A=C(T_2-\alpha)-D\beta,\hspace{4mm} -C\beta+D(T_1-\alpha)=0.
$$
Expressing $C=D(T_1-\alpha)(\beta)^{-1}$ from the second equation and plugging it in to the first, we get that
$A=D(t_1\beta)^{-1}H_1$. The case of $\psi_2$ is similar.

These two propositions allow us to write our hierarchy as a system of rank one difference equations of degree
two. 

\begin{proposition} The discrete modified Novikov-Veselov hierarchy (\ref{dmNV}) is equivalent to either of
the following two systems of equations
\begin{equation}
\frac{\partial}{\partial \tau_s^i}H_1+H_1A_{s,1}^i\equiv 0 \mbox{ mod } H_1,
\label{dmNV2}
\end{equation}
\begin{equation}
\frac{\partial}{\partial \tau_s^i}H_2+H_2A_{s,2}^i\equiv 0 \mbox{ mod } H_2.
\label{dmNV3}
\end{equation}
\end{proposition}
We now use this approach to construct the equations corresponding to times $\tau_1^1$, $\tau_1^2$, $\tau_2^1$ and $\tau_2^2$.

The functions $\psi_1$ and $\psi_2$ have the following power series expansions at the marked points $P_i^{\pm}$, where by $k$ we denote the appropriate local coordinate $k_i^{\pm}$:
\begin{equation}
\begin{array}{c}
\psi_1(n,m,\tau;P)=k^{\pm n}\exp\left(\pm\displaystyle\sum_{\sigma=1}^{\infty}\tau_\sigma^1k^{\sigma}\right)\cdot
\left(\displaystyle\sum_{\alpha=0}^{\infty}\xi_{1,\alpha}^{\pm}(n,m,\tau)k^{-\alpha}\right)\mbox{ as }P\rightarrow P_1^{\pm},\\
\psi_1(n,m,\tau;P)=k^{\pm m}\exp\left(\pm\displaystyle\sum_{\sigma=1}^{\infty}\tau_\sigma^2k^{\sigma}\right)\cdot
\left(\displaystyle\sum_{\alpha=0}^{\infty}\xi_{2,\alpha}^{\pm}(n,m,\tau)k^{-\alpha}\right)\mbox{ as }P\rightarrow P_2^{\pm},\\
\psi_2(n,m,\tau;P)=k^{\pm n}\exp\left(\pm\displaystyle\sum_{\sigma=1}^{\infty}\tau_\sigma^1k^{\sigma}\right)\cdot
\left(\displaystyle\sum_{\alpha=0}^{\infty}\chi_{1,\alpha}^{\pm}(n,m,\tau)k^{-\alpha}\right)\mbox{ as }P\rightarrow P_1^{\pm},\\
\psi_2(n,m,\tau;P)=k^{\pm m}\exp\left(\pm\displaystyle\sum_{\sigma=1}^{\infty}\tau_\sigma^2k^{\sigma}\right)\cdot
\left(\displaystyle\sum_{\alpha=0}^{\infty}\chi_{2,\alpha}^{\pm}(n,m,\tau)k^{-\alpha}\right)\mbox{ as }P\rightarrow P_2^{\pm},\end{array}
\end{equation} 
where the $\xi_{i,s}^{\pm}(n,m,\tau)$ and $\chi_{i,s}^{\pm}(n,m,\tau)$ are analytic functions in the variables
$\tau$, and $\xi_{2,0}^+=0$, $\chi_{1,0}^+=0$. To make our notation consistent with (\ref{psi1expansionsnotime})-(\ref{psi2expansionsnotime}), we denote
\begin{equation}
a_i^{\pm}=\xi_{i,0}^{\pm},\hspace{4mm}b_i^{\pm}=\chi_{i,0}^{\pm}
\end{equation}
\begin{equation}
c_i^{\pm}=\xi_{i,1}^{\pm},\hspace{4mm}d_i^{\pm}=\chi_{i,1}^{\pm}
\end{equation}
Plugging these expressions into (\ref{Diractime}), we see that these coefficients satisfy the following system of equations:
\begin{equation}
t_2\xi_{1,\alpha}^{\pm}=\alpha\xi_{1,\alpha}^{\pm}+\beta\chi_{1,\alpha}^{\pm}
\label{main1}
\end{equation}
\begin{equation}
t_2\xi_{2,\alpha\pm 1}^{\pm}=\alpha\xi_{2,\alpha}^{\pm}+\beta\chi_{2,\alpha}^{\pm}
\label{main2}
\end{equation}
\begin{equation}
t_1\chi_{1,\alpha\pm 1}^{\pm}=\beta\xi_{1,\alpha}^{\pm}+\alpha\chi_{1,\alpha}^{\pm}
\label{main3}
\end{equation}
\begin{equation}
t_1\chi_{2,\alpha}^{\pm}=\beta\xi_{2,\alpha}^{\pm}+\alpha\chi_{2,\alpha}^{\pm}
\label{main4}
\end{equation}
Also, since the functions $\psi_1$ and $\psi_2$ satisfy the normalization conditions (\ref{norm1})-(\ref{norm3}), we also
have
\begin{equation}
a_1^+a_1^-=1,\hspace{4mm}b_2^+b_2^-=1.
\label{normtime}
\end{equation}


We now derive the dmNV equation corresponding to time $\tau_1^1$ using its equivalent
form (\ref{dmNV2}). Let $\dot{f}$ denote
differentiation by $\tau_1^1$. We denote $A_{1,1}^1=AT_1+BT_1^{-1}+C$ and $H_1=T_1T_2+xT_1+yT_2+z$. The equation in
time $\tau_1^1$ has the form
\begin{equation}
-\dot{x}T_1-\dot{y}T_2-\dot{z}\equiv (T_1T_2+xT_1+yT_2+z)(AT_1+BT_1^{-1}+C)\mbox{ mod }H_1.
\label{mdNV11}
\end{equation}
First, we express all of the coefficients of the above equation in terms of the variables $a_1^+$, $b_2^+$, $\alpha$
and $\beta$. The coefficients $x$, $y$, $z$ of $H_1$ were found above in Proposition 4.1:
\begin{equation}
x=-t_1\alpha,\hspace{4mm}y=-\frac{\alpha(t_1\beta)}{\beta},\hspace{4mm}z=\frac{t_1\beta}{\beta}.
\end{equation}
To calculate the coefficients of the operator $A_{1,1}^1$, we use the method of Proposition 3.3. Comparing singularities,
we see that if
\begin{equation}
A=f^1_{1,1,1}=\frac{a_1^+}{t_1a_1^+},\hspace{4mm}B=f^1_{1,1,-1}=-\frac{a_1^-}{t_1^{-1}a_1^-}=-\frac{t_1^{-1}a_1^+}{a_1^+},
\label{A1}
\end{equation}
then the functions $\psi_1$ and $\dot{\psi_1}-AT_1\psi_1-BT_1^{-1}\psi$ are proportional. Hence we can determine
the third coefficient $C=f^1_{1,1,0}$ by comparing these two functions at either $P_2^+$ or $P_2^-$, which gives us
two alternative expressions:
\begin{equation}
C=f^1_{1,1,0}=
\frac{1}{c_2^+}\left(\frac{\partial c_2^+}
{\partial \tau_1^1}-\frac{a_1^+}{t_1a_1^+}t_1c_2^++\frac{a_1^-}{t_1^{-1}a_1^-}t_1^{-1}c_2^+\right)=
\frac{1}{a_2^-}\left(\frac{\partial a^-_2}{\partial \tau_1^1}-\frac{a_1^+}{t_1a_1^+}t_1a_2^-+
\frac{a_1^-}{t_1^{-1}a_1^-}t_1^{-1}a_2^-\right).
\label{A1second}
\end{equation}
We first these expressions by removing the coefficients $a_2^-$ and $c_2^+$. From the system (\ref{main1}-\ref{main4})
we get that $c_2^+=(t_2^{-1}\beta)(t_2^{-1}b_2^+)$ and 
$a_2^-=-\beta /(\alpha b_2^+)$. Using $t_1b_2^+=\alpha b_2^+$, the first expression becomes
$$
C=\frac{t_2^{-1}\dot{\beta}}{t_2^{-1}\beta}+\frac{t_2^{-1}\dot{b}_2^+}{t_2^{-1}b_2^+}
-\frac{a_1^+}{t_1a_1^+}\frac{(t_2^{-1}\alpha)(t_1t_2^{-1}\beta)}{t_2^{-1}\beta}+
\frac{t_1^{-1}a_1^+}{a_1^+}\frac{t_1^{-1}t_2^{-1}\beta}{(t_2^{-1}\beta)(t_1^{-1}t_2^{-1}\alpha)}
$$
and the second expression becomes
$$
C=f^1_{1,1,0}=\frac{\dot{\beta}}{\beta}-\frac{\dot{\alpha}}{\alpha}-\frac{\dot{b}_2^+}{b_2^+}-
\frac{a_1^+}{t_1a_1^+}\frac{t_1\beta}{\beta(t_1\alpha)}+\frac{t_1^{-1}a_1^+}{a_1^+}\frac{\alpha(t_1^{-1}\beta)}{\beta}.
$$

Expanding the right hand side of (\ref{mdNV11}), we get
$$
H_1A^1_{1,1}=(t_1t_2A)T_1^2T_2+x(t_1A)T_1^2+\left[t_1t_2C+y(t_2A)\right]T_1T_2+\left[x(t_1C)+zA\right]T_1+
$$
$$
+\left[t_1t_2B+y(t_2C)\right]T_2+x(t_1B)+zC+y(t_2B)T_1^{-1}T_2+zBT_1^{-1}.
$$
This expression is a Laurent polynomial in $T_1$ and $T_2$ whose
terms have degrees $i$ and $j$ in $T_1$ and $T_2$, respectively, where $i=-1,0,1,2$ and $j=0,1$. We need to express it as a left
multiple of $H_1$ plus an operator containing terms of degrees $(0,0)$, $(0,1)$ and $(1,0)$. First, to cancel the term
containing $T_1^2T_2$, we subtract the following left multiple of $H_1$:
$$
(t_1t_2A)T_1H_1=(t_1t_2A)T_1^2T_2+(t_1t_2A)(t_1x)T_1^2+(t_1t_2A)(t_1y)T_1T_2+(t_1t_2A)(t_1z)T_1.
$$
Using (\ref{Sch}), (\ref{A1}) and (\ref{main1}), we see that the coefficient in front of $T_1^2$ in this difference vanishes:
$$
x(t_1A)-(t_1x)(t_1t_2A)=-(t_1\alpha)\frac{t_1a_1^+}{t_1^2a_1^+}+(t_1^2\alpha)\frac{t_1t_2a_1^+}{t_1^2t_2a_1^+}=0.
$$
Similarly, to cancel the term containing $T_1^{-1}T_2$, we subtract 
$$
y(t_2B)T_1^{-1}y^{-1}H_1=\frac{y(t_2B)}{t_1^{-1}y}T_2+\frac{y(t_2B)}{t_1^{-1}y}(t_1^{-1}x)+
y(t_2B)T_1^{-1}T_2+\frac{y(t_2B)}{t_1^{-1}y}(t_1^{-1}z)T_1^{-1},
$$
and using (\ref{Sch}), (\ref{A1}), (\ref{main1}) and the relation (\ref{normtime}), we show that the coefficient in front
of $T_1^{-1}$ vanishes:
$$
zB-\frac{y(t_2B)}{t_1^{-1}y}(t_1^{-1}z)=0.
$$
Hence, we see that
$$
H_1A^1_{1,1}\equiv\left[t_1t_2C+y(t_2A)-(t_1t_2A)(t_1y)\right]T_1T_2+\left[x(t_1C)+zA-(t_1t_2A)(t_1z)\right]T_1+
$$
$$
+\left[t_1t_2B+ y(t_2C)-\frac{y(t_2B)}{t_1^{-1}y}\right]T_2+x(t_1B)+zC-\frac{y(t_2B)}{t_1^{-1}y}(t_1^{-1}x)\mbox{ mod }H_1.
$$
Finally, to obtain the evolution equation, we subtract $\left[t_1t_2C+y(t_2A)-(t_1t_2A)(t_1y)\right]H_1$ from the right
hand side of the equation, and obtain the following equations:
\begin{equation}
-\dot{x}=x(t_1C)+zA-(t_1t_2A)(t_1z)-x\left[t_1t_2C+y(t_2A)-(t_1t_2A)(t_1y)\right],
\end{equation}
\begin{equation}
-\dot{y}=t_1t_2B+y(t_2C)-\frac{y(t_2B)}{t_1^{-1}y}-y\left[t_1t_2C+y(t_2A)-(t_1t_2A)(t_1y)\right],
\end{equation}
\begin{equation}
-\dot{z}=x(t_1B)+zC-\frac{y(t_2B)}{t_1^{-1}y}(t_1^{-1}x)-\left[t_1t_2C+y(t_2A)-(t_1t_2A)(t_1y)\right].
\end{equation}
Since the coefficients $x$, $y$, $z$ of $H$ are expressed in terms of $\alpha$ and $\beta$, which are in turn related
by the equation $\alpha^2-\beta^2=1$, it is sufficient to find one of the derivatives, for
example $\dot{x}$. Expanding the expression for $\dot{x}$ and using the expressions for the coefficients $x$, $y$, $z$ and $A$, $B$, $C$ obtained above (using the first expression
for $C$ in $t_1t_2C$ and using the second one in $t_1C$), we obtain the following equation
\begin{equation}
\frac{t_1\dot{b}_2^+}{t_1b_2^+}=\frac{a_1^+}{t_1a_1^+}\frac{\beta(t_1\beta)}{t_1\alpha},
\label{pretime1}
\end{equation}
which is the first equation of the dmNV hierarchy.

It seems natural to replace the variables $a_1^+$ and $b_2^+$ with their logarithms, i.e. to introduce
 new variables $a_1^+=e^{\varphi}$ and $b_2^+=e^{\psi}$. Since $\alpha=t_2a_1^+/a_1^+=t_1b_2^+/b_2^+$, these
variables are related by the equation
\begin{equation}
t_2\varphi-\varphi=t_1\psi-\psi.
\end{equation}
Writing the evolution equation (\ref{pretime1}) in terms of these new variables, we get
\begin{equation}
\frac{\partial \psi}{\partial \tau_1^1}=\sqrt{\left(e^{2t_1^{-1}t_2 \varphi} -e^{2t_1^{-1}\varphi}\right)
\left(e^{-2\varphi}-e^{-2t_2\varphi}\right)}
\label{t11final}
\end{equation}

To derive the evolution equation for time $\tau_1^2$, we use its equivalent form (\ref{dmNV3}). The calculations
in this case are identical to those performed above. In fact, since our problem is symmetric with respect to exchanging the marked points $P_1^{\pm}$ and $P_2^{\pm}$, we can obtained the desired equation simply by exchanging the functions $a_1^+$ and
$b_2^+$ and simultaneously exchanging the shift operators $t_1$ and $t_2$ in the
evolution equation in time $\tau_1^1$ (\ref{t11final}). This
gives us the following equation:
\begin{equation}
\frac{t_2 \dot{a}_1^+}{t_2 a_1^+}=\frac{\beta (t_2\beta)}{t_2\alpha}\frac{b_2^+}{t_2b_2^+}.
\end{equation}
In terms of the logarithmic variables, this equation reads
\begin{equation}
\frac{\partial \varphi}{\partial \tau_1^2}=\sqrt{\left(e^{2t_1t_2^{-1}\psi}-e^{2t_2^{-1}\psi}\right)
\left(e^{-2\psi}-e^{-2t_1\psi}\right)}
\label{t12final}
\end{equation}

The derivation of the equations for times $\tau_2^1$ and $\tau_2^2$ involves similar calculations. For time $\tau_2^1$, we use
the equivalent form (\ref{dmNV2}):
\begin{equation}
-\frac{\partial H_1}{\partial \tau_2^1}=H_1A_{2,1}^1\mbox{ mod }H_1.
\end{equation}
Here $A_{2,1}$ is a Laurent polynomial in $T_1$ with terms of degree $-2$ to $3$. As above, we successively subtract appropriate left multiples of $H_1$ to cancel the terms containing $T_1^iT_2$ for $i=3,-2,2,-1$. At every step, the
corresponding $T_1^i$ term vanishes. Finally, canceling the $T_1T_2$ term gives us the following equation:
\begin{equation}
\frac{t_1\dot{b}_2^+}{t_1b_2^+}=\frac{\beta(t_1\beta)}{t_1\alpha}\frac{1}{t_1a_1^+}c_1^+-
\frac{\beta(t_1\beta)}{t_1\alpha}\frac{a_1^+}{(t_1a_1^+)(t_1^2a_1^+)}t_1^2c_1^++
\frac{\beta(t_1^2\beta)}{(t_1\alpha)(t_1^2\alpha)}\frac{a_1^+}{t_1^2a_1^+}+
\frac{\alpha(t_1^{-1}\beta)(t_1\beta)}{t_1\alpha}\frac{t_1^{-1}a_1^+}{t_1a_1^+},
\label{t21final}
\end{equation}
where the functions $a_1^+$, $b_2^+$, $c_1^+$, $\alpha$ and $\beta$ in the equation satisfy the following relations:
\begin{equation}
\alpha=\frac{t_2 a_1^+}{a_1^+}=\frac{t_1b_2^+}{b_2^+},\hspace{4mm}\alpha^2-\beta^2=1,\hspace{4mm}
t_2c_1^+=\alpha c_1^++\beta(t_1^{-1}\beta)(t_1^{-1}a_1^+).
\end{equation}

\section{Explicit formulas}

\setcounter{equation}{0}

In this section we give explicit formulas for the functions $\psi_i(n,m,\tau;P)$ in terms of the theta-functions
of the surface $X$. Choose a basis
$a_j$, $b_j$, $j=1,\ldots,g$ of $H_1(X,\mathbb{Z})$ with canonical intersection form, i.e. such that $a_j\circ a_k=0$, 
$b_j\circ b_k=0$, $a_j\circ b_k=\delta_{jk}$. Let $B$ be the period
matrix of the curve $X$ with respect to this basis. Let $\Omega^1_1$ and
$\Omega^1_2$ denote Abelian differentials of the third kind with poles at $P_1^{\pm}$ and $P_2^{\pm}$:
$$
\Omega^1_i=d( k_i^{\pm})^{-1}\left(\mp k_i^{\pm}+O(1)\right)\mbox{ as }P\rightarrow P_i^{\pm}
$$
which are normalized to have zero periods over the $a$-cycles. Let $\Omega^s_{i}$ denote Abelian differentials of the second
kind with poles at $P_i^{\pm}$ and principal parts
$$
\Omega^s_i=d (k_i^{\pm})^{-1}\left(\mp s(k_i^{\pm})^{s+1}+O(1)\right)\mbox{ as }P\rightarrow P_i^{\pm},
$$
and with zero $a$-periods, and which are odd with respect to the involution $\sigma$. It is a standard fact that these
differentials exist and are unique. Let $U^1_i$ and $U^k_i$ denote the vectors of the $b$-periods of these differentials:
$$
(U^1_i)_j=\frac{1}{2\pi i}\displaystyle\oint_{b_j}\Omega^1_i,\hspace{4mm}
(U^s_i)_j=\frac{1}{2\pi i}\displaystyle\oint_{b_j}\Omega^s_i.
$$
Choose a base point $P_0\in X$ away from the marked points $P_i^{\pm}$ and the divisor $D$, and let
$A:X\rightarrow J(X)$ denote the Abel map with base point $P_0$, where $J(X)$ is the Jacobian
variety of $X$. Let $\theta(z|B)$ denote the theta function of $J(X)$ for $z\in\mathbb{C}^g$.
Introduce the functions
$$
r_1(P)=\frac{\theta(A(P)-A(P_2^+)-\sum_{i=2}^gA(P_i)-K|B)\theta(A(P)-\sum_{i=1}^{g+1}A(P_i)+A(P_2^+)-K|B)}
{\theta(A(P)-\sum_{i=1}^gA(P_i)-K|B)\theta(A(P)-\sum_{i=2}^{g+1}A(P_i)-K|B)},
$$
$$
r_2(P)=\frac{\theta(A(P)-A(P_1^+)-\sum_{i=2}^gA(P_i)-K|B)\theta(A(P)-\sum_{i=1}^{g+1}A(P_i)+A(P_1^+)-K|B)}
{\theta(A(P)-\sum_{i=1}^gA(P_i)-K|B)\theta(A(P)-\sum_{i=2}^{g+1}A(P_i)-K|B)}.
$$
By construction, these are meromorphic functions on $X$ whose pole divisor is $D=\sum_{i=1}^{g+1}P_i$ and
whose zero divisors are $P_2^++D_1$ and $P_1^++D_2$, respectively, where $D_1$ and $D_2$ are some divisors of degree $g$.

We define the functions $\psi_1$ and $\psi_2$ by the following formulas:
\begin{equation}
\psi_i(n,m,\tau;P)=r_i(P)C_i(n,m,\tau)F_i(n,m,\tau;P)
\exp\left[n\displaystyle\int_{P_0}^P\Omega^1_1+m\displaystyle\int_{P_0}^P\Omega^1_2+\displaystyle
\sum_{s=1}^{\infty}\displaystyle\sum_{i=1}^2
\tau^i_s\displaystyle\int_{P_0}^P\Omega^s_{i}\right],
\label{theta-formula}
\end{equation}
where the function $F(n,m,\tau;P)$ is defined as
$$
F_i(n,m,\tau;P)=
\frac{\theta\left(A(P)-A(D_i)+nU^1_1+mU^1_2+\displaystyle\sum_{s=1}^{\infty}\displaystyle\sum_{i=1}^2
\tau^i_s U^s_i\right)}{\theta\left(A(P)-A(D_i)-K\right)}
$$
and the path of integration in the exponent is the same as in the Abel map in $F_i$. By construction, these are single-valued
functions on the curve $X$, having the required meromorphic and exponential singularities at the marked points, and having
pole divisor $D$ away from the marked points. 

The constants $C_i(n,m,\tau)$ are determined by the normalization 
conditions (\ref{norm1})-(\ref{norm3}). Choose paths of integration $\gamma_i:[0,1]\rightarrow X$ from $P_0$ to $P_i^+$ and
a path $\gamma$ from $P_0$ to $\sigma(P_0)$. We assume that the integration path in $\psi_i(P)$ is $\gamma_i$
and that the path in $\psi_i(\sigma(P))$ is $\gamma$ followed by the image of $\gamma_i$ under $\sigma$. Writing
out the expression for $\psi_i(P)\psi_i(\sigma(P))$ using (\ref{theta-formula}), we see that we need
to choose the constants $C_i(n,m,\tau)$ as follows:
\begin{equation}
\frac{1}{C_i(n,m,\tau)^2}=r_i(P_i^+)r_i(P_i^-)F_i(n,m,\tau;P_i^+)F_i(n,m,\tau;P_i^-)
\exp\left[nI_i^1+mI_i^2+\displaystyle
\sum_{s=1}^{\infty}\displaystyle\sum_{i=1}^2
\tau^i_s\displaystyle\int_{\gamma}\Omega^s_{i}\right]
\end{equation}
where the path of integration in the $F_i(n,m,\tau;P_i^-)$ factor is $\gamma$ followed by $\sigma(\gamma_i)$,
and the constants $I_i^1$ and $I_i^2$ are the principal values of the integrals of $\Omega_1^1$ and $\Omega_2^1$
along the path $-\gamma_i+\gamma+\sigma(\gamma_i)$:
\begin{equation}
I_i^k=\displaystyle\lim_{t\rightarrow 1}\left(\displaystyle\int_{\gamma(0)}^{\gamma(t)}\Omega_k^1+
\displaystyle\int_{\gamma}\Omega_k^1+
\displaystyle\int_{\sigma(\gamma(0))}^{\sigma(\gamma(t))}\Omega_k^1\right),\hspace{4mm}k=1,2.
\end{equation}

 Finally, we choose the signs of $C_i(n,m,\tau)$ in such a way that the functions $\psi_i$ satisfy the equation (\ref{norm3}).

\section{Acknowledgments}

The author would like to sincerely thank I. M. Krichever for suggesting the problem, for many useful and interesting
discussions, and for pointing out errors in an early draft of the work.

\end{document}